\documentclass[10pt,journal,compsoc]{IEEEtran}

\usepackage{graphicx}
\usepackage{amsmath,amssymb}

\usepackage{tabularx}
\usepackage{multirow}
\usepackage{dcolumn}
\usepackage{booktabs}
\usepackage{enumerate}

\usepackage{subfigure}
\graphicspath{{./images/}}

\usepackage{pgfplots}
\pgfplotsset{compat=1.12}
\usetikzlibrary{patterns}
\usetikzlibrary{spy}

\usepackage{tikz}

\usepackage{algpseudocode}
\usepackage{algorithm}

\algnewcommand\algorithmicforeach{\textbf{for each}}
\algdef{S}[FOR]{ForEach}[1]{\algorithmicforeach\ #1\ \algorithmicdo}

\usepackage{amsthm}

\usepackage{nomencl}
\makenomenclature

\usepackage{etoolbox}
\renewcommand\nomgroup[1]{%
  \item[\bfseries
  \ifstrequal{#1}{A}{Acronyms}{%
  \ifstrequal{#1}{S}{Symbols}{%
  \ifstrequal{#1}{U}{Subscripts and Superscripts}{}}}%
]}

\usepackage{hyperref}
\hypersetup{
  urlcolor   = blue,
  colorlinks = true,
}

\usepackage[normalem]{ulem}

\ifCLASSOPTIONcompsoc
  \usepackage[nocompress]{cite}
\else
  \usepackage{cite}
\fi

\begin{document}

\title{Audiovisual Saliency Prediction in Uncategorized Video Sequences based on Audio-Video Correlation}

\author{Maryam Qamar Butt, Anis Ur Rahman \\

\thanks{Maryam Q. Butt and Anis Ur Rahman are with School of Electrical Engineering and Computer Science (SEECS), National University of Sciences and Technology (NUST), Islamabad, Pakistan. 
Corresponding Author: Anis U. Rahman, e-mail: anis.rahman@seecs.edu.pk}
%
}
\maketitle

\begin{abstract}
Substantial research has been done in saliency modeling to develop intelligent machines that can perceive and interpret their surroundings. But existing models treat videos as merely image sequences excluding any audio information, unable to cope with inherently varying content. Based on the hypothesis that an audiovisual saliency model will be an improvement over traditional saliency models for natural uncategorized videos, this work aims to provide a generic audio/video saliency model augmenting a visual saliency map with an audio saliency map computed by synchronizing low-level audio and visual features. The proposed model was evaluated using different criteria against eye fixations data for a publicly available DIEM video dataset. The results show that the model outperformed two state-of-the-art visual saliency models.
\end{abstract}


\IEEEpeerreviewmaketitle


\section{Introduction}
\label{sec:intro}




Though a lot of research has been done in the general field of unimodal saliency models for both images and videos, no substantial contributions exist for bimodal models. Of more consequence is the lack of a model for computation of audiovisual saliency in complex video sequences. Existing literature for audio-video saliency modeling is scarce and often targets a specific class of videos~\cite{rapantzikos2009spatiotemporal,coutrot2016multimodal,ould2016audiovisual}. Therefore, an extended saliency model to predict salient regions in complex videos with different sound classes is required.

Many existing saliency algorithms are designed for images~\cite{bruce2005saliency,le2006coherent,murray2011saliency} using visual cues such as color, intensity, orientation etc., while other models~\cite{cerf2008predicting,judd2009learning,marat2013improving} take social cues like faces into account resulting in more accurate eye movement predictions. Spatiotemporal saliency models~\cite{le2007predicting,guo2010novel,marat2009modelling} usually incorporate temporal cues like motion but ignore the effect of audio stimuli--an integral component of video content--on human gaze. Subsequently, such models are classified as unimodal models~\cite{borji2013state} where only visual stimuli are used.

Interestingly, the effect of audio stimuli is relevant to human eye movements. In~\cite{quigley2008audio} the authors find eye movements to be spatially biased towards the source of audio using an eye tracking experiment on images with spatially localized sound sources in three conditions: auditory (A), visual (V) and audio-visual (AV). Moreover, another study~\cite{song2013different} analyzed the effects of different type of sounds on human gaze involving an experiment with thirteen sound classes under audio-visual and visual conditions. The sound classes are further clustered into on-screen with one sound source, on-screen with more than one sound source, and off-screen sound source. The results show that human speech, singer(s) and human noise (on-screen sound source clusters) highly affect gaze and, more importantly, linked audio-visual stimuli has a greater effect than unsynchronized audio-visual events.

The focus of this work is to propose a generic audio-visual saliency model for complex video sequences. The work differs from previous research~\cite{rapantzikos2009spatiotemporal,coutrot2016multimodal,ould2016audiovisual} in that it does not restrict input videos from a certain category. To accomplish that an audio source localization method was used to relate an audio signal with an object in the video frames in a rank correlation space. The proposed model was evaluated against eye fixations ground truth from DIEM dataset.



The original contribution of this study is as follows:
\begin{enumerate}
\item Propose an audio-visual saliency model for complex scenes that, unlike existing literature, does not restrain videos to any specific category.
\item Present and analyze the results of experimental evaluation on a publicly available dataset to examine how our proposed saliency model compares to two state-of-the-art audio-visual saliency models.
\end{enumerate}


The remainder of the paper is organized as follows: Section~\ref{sec:intro} narrates background knowledge of saliency modeling and identifies the novel contribution of this work. Section~\ref{sec:rel} provides a detailed review of state-of-the-art literature while Section~\ref{sec:ps} describes the proposed solution. Section~\ref{sec:mm} summarizes the implementation details as well as outlines the properties of video sequences used for experimentation. This section also explains the different saliency evaluation metrics. Section~\ref{sec:result} presents our results followed by a discussion in Section~\ref{sec:disc}. Section~\ref{sec:conclusion} summarizes our findings and concludes with future perspectives.




\section{Related work}
\label{sec:rel}

Unimodal saliency models use only one type of sensory stimulus as input, some visual cues including color, intensity and orientation features~\mbox{\cite{murray2011saliency,borji2011cost,avraham2010esaliency}}. Other biologically-inspired models~\mbox{\cite{marat2007video,marat2009modelling}} exploit spatial contrast and motion, and simulate interactions between neurons using excitation and inhibition mechanisms. While others~\mbox{\cite{liu2014superpixel,liu2016saliency}} propagate spatial/ temporal saliency using multiscale color and motion histograms as features. In~\cite{liu2014superpixel} pixel-level spatiotemporal saliency is computed from spatial and temporal saliencies via interaction and selection driven from superpixel-level saliency. In~\cite{liu2016saliency} temporal saliency is propagated forward and backward via inter-frame similarity matrices and graph-based motion saliency, whereas spatial saliency is propagated over a frame using temporal saliency and intra-frame similarity matrices. In most of these models conspicuity maps are constructed using a variety of approaches with different visual features that are later integrated together to get a final saliency map.

Based on the fact that eyes are the most important sensory organs that provide much of the information around humans, many state-of-the-art visual models~\cite{liu2014superpixel,liu2016saliency} aim at saliency computation for complex dynamic scenes. But such unimodal models tend to overlook other influential social cues like faces in social interaction scenes, and hence exhibit lower predictability~\cite{birmingham2009saliency,tatler2011eye}. Moreover, social scenes involve a lot more sensory signals influencing eye movements spatially such as auditory information including voice tone, music, etc, and different kinds of sounds affect eye fixations differently~\cite{quigley2008audio,song2013different}. Thus, there is a need for a bimodal saliency model incorporating both visual and audio information channels.

Rapantzikos et al.~\mbox{\cite{rapantzikos2007audio}} proposed an audio-visual saliency model for movie summarization. The visual saliency map is constructed using traditional features such as intensity, color and motion, and simulating feature competition as energy minimization via gradient descent. This map is thresholded and averaged per frame to compute a $1D$ visual saliency curve. While maximum average Teager energy, mean instant amplitude and mean instant frequency, are extracted as audio features by applying Teager Kaiser energy operator and energy separation algorithm on the audio signal. The resulting feature vector is normalized to a range $[0, 1]$ followed by weighted fusion to get an audio saliency curve. The final audio-visual saliency curve is a weighted linear combination of audio and visual saliency curves. The local maxima feature of audio-visual saliency curve is used for key-frame selection. The experiments are conducted on movie database of A.U.T.H but no comparison and evaluation is given.

Coutrot and Guyader~\cite{coutrot2014audiovisual} proposed an audiovisual saliency model for natural conversation scenes; a linear combination of low-level saliency, face map, and center bias. Low-level saliency map is constructed via Marat's spatiotemporal saliency model~\cite{marat2009modelling}. While for face map construction a  speaker diarization algorithm is proposed that uses motion activity of faces and 26 Mel-frequency cepstral coefficients (MFCCs) as visual and audio features respectively. Center bias is a time-independent 2D Gaussian function centered on the screen. The three maps are linearly combined into final audiovisual saliency map using expectation maximization to determine the weight for each. The resulting model performs better compared to the same model without speaking and mute face differentiation. However, the target video dataset belongs to a limited category: conversation scenes only.

Sidaty et al.~\cite{ould2016audiovisual} proposed an audiovisual saliency model for teleconferencing and conversational videos. Three best performing models on target database i.e. Itti et al~\cite{itti1998model}, Harel et al.~\cite{harel2006graph} and Tavakoli et al.~\cite{tavakoli2011fast} are selected as spatial models. Acoustic energy is computed per frame and block matching algorithm is used to construct an audio map using the face stream of video. Then peak matching is used for audio-visual synchronization. Five fusion schemes are used to get a final map. Experiments performed on XLIMedia database created by the authors showed that the proposed model performed better compared to spatial models. Again the limitation of this work is that it only targets conferencing and conversational videos.

All in all, one of the major limitation of the aforementioned visual models is that they treat videos as a mute sequence of images and ignore any influence of audio stimuli. This results in inaccurate predictions where sound guides eye movement. Furthermore, another limitation of literature is the absence of an audiovisual model for complex dynamic scenes; that is, many of the state-of-the-art models restrict the dataset used to only one specific category, for instance, conversational videos. This limits the models' performance when dealing with videos containing different sound classes.


\section{Proposed Solution}
\label{sec:ps}
\begin{figure*}[!ht]
\centering
\includegraphics[width=1.0\textwidth]{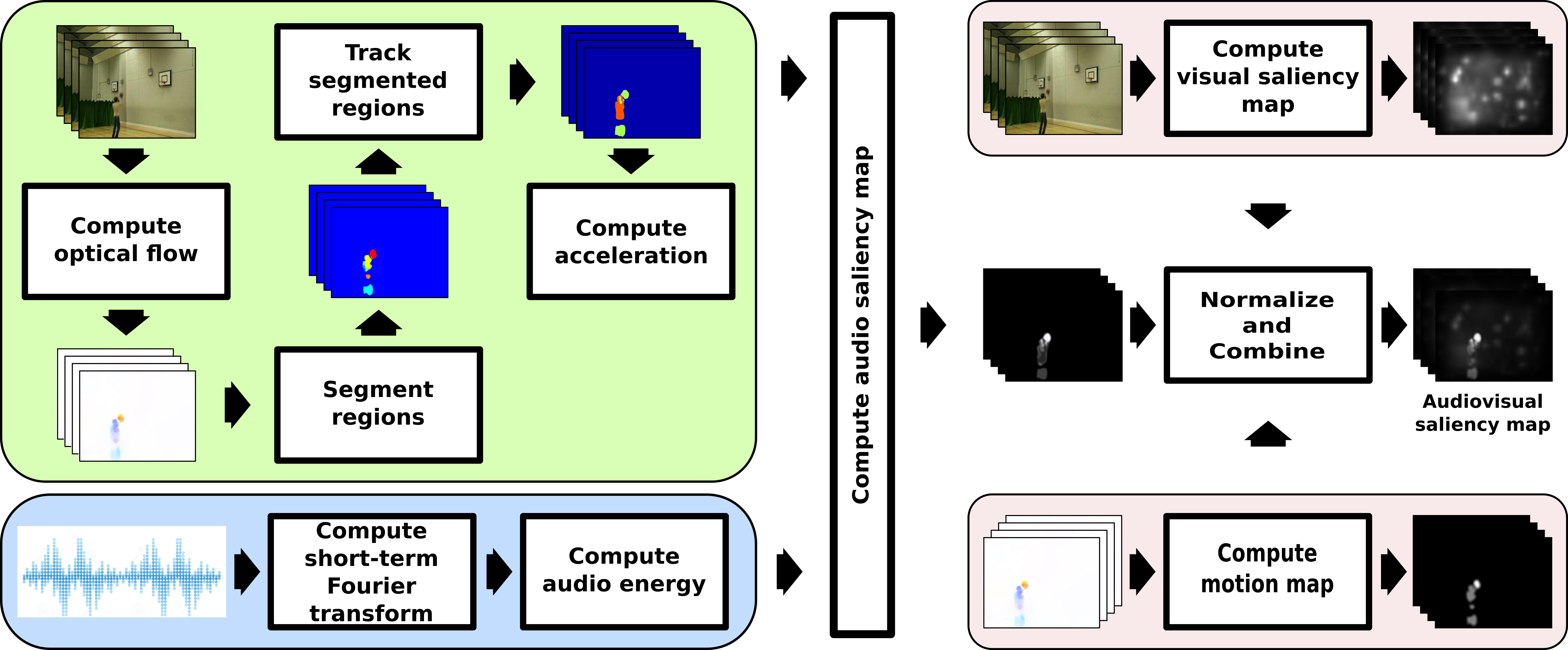}
\caption{Architecture of proposed solution}
\label{fig:proposed_framework}
\end{figure*}

This section explains the proposed solution for audio-visual saliency computation for videos. The framework consists of five major stages as illustrated in Figure~\ref{fig:proposed_framework}. The first stage is the extraction of audio energy descriptors and object motion descriptors per frame using audio and visual stimuli as separate channels. The next stage computes an audio saliency map using these descriptors. In parallel, another stage computes visual saliency map and motion map. The former using low-level features while the latter from a color-coded optical flow similar to one done for the audio maps. The last stage normalizes and combines all these maps into a unified audiovisual saliency map.

\subsection{Feature Extraction}

In this stage, we extracted visual and acoustic features from a given input video. The stage comprised two phases of feature extraction, one for audio features and the other for visual features.

\subsubsection{Audio Feature Extraction}

The step outputs an audio energy descriptor $a(t)$ extracted from an audio signal featuring changing patterns of an audio signal strength. Note that the signal was obtained with the same temporal resolution as the video frames. Hence, the signal was first segmented into frames according to the frame rate of video so-that each audio frame corresponds to a video frame. Using short-term Fourier transform (STFT), this framed signal was transformed into a time-frequency domain to get a spectrogram of the signal at each frame. The descriptor $a(t)$ was computed by the integration of the resultant spectrogram at any given frame over all frequencies using,
\begin{align*}
a(t)=\int_0^{\infty} \int_0^T f(t^\prime) W(t^\prime-t)e^{-j2 \pi ft^\prime} dt^\prime df
\end{align*}
where the windowing function $W(t)$ is defined so that neighboring windows overlap by $50\%$. The final descriptor was post-processed using a $1D$ Gaussian kernel.

\subsubsection{Visual Feature Extraction}
\label{sec:visualfeatures}

Based on the assumption that a moving object is a prime candidate to be an audio signal source, acceleration per frame of all moving objects in a given input video was computed as motion descriptor. First, the moving objects were segmented per frame using optical flow estimation and tracked along with all frames via color histograms of the regions in HSV color space. The process is described as follows:

\begin{enumerate}

\item \textbf{Optical flow computation.} 
The method proposed by~\cite{chang2013superpixel} was used to compute dense optical flow and corresponding color-coded optical flow images per video frame. The method used apparent motion of each pixel to compute forward and backward optical flows where the former depicts the motion of pixels of frame $t$ with reference to frame $t+1$ and the latter was the motion of pixels of frame $t$ with respect to frame $t-1$. The resulting flows were averaged out to get a mean optical flow per frame, later used to compute an audio saliency map. 

\item \textbf{Frame segmentation.} 
The color-coded mean optical flow per frame was used as input for the segmentation step. Mean shift, a nonparametric clustering algorithm was applied to segment input image in LUV color space. The oversegmented result of the step was followed by a simple region merging technique based on DeltaE, a color difference score, to merge the closely similar regions. Regions smaller than 200 pixels were filtered as noise and insignificant regions.

\item \textbf{Region tracking.}
Once individual frames were segmented, a number of tracks were initialized in the first frame using the segmented regions' location and appearance features. All regions in following segmented input frames were either assigned to an existing track or initialized to a new track based on its location and appearance similarities. The location similarity $d_E$ was computed by Euclidean distance between the centroid of a new region $C_n$ and that of an existing track $C_e$ using,
\begin{align*}
d_E=\sqrt{(C_n(x,y)-C_e(x,y))^2}
\end{align*}
This resulted in a list of candidate tracks similar to the region under consideration for assignment within a specified search radius \($r$\). For appearance similarity, $AS$ LUV histograms of existing candidate tracks $H_e$ were compared to the new region's histogram $H_n$ using cosine similarity $cos\theta$ as,
\begin{align*}
cos\theta=\frac{H_n \cdot H_e}{\left \| H_n \right \| \left \| H_e \right \|}
\end{align*}
The region $C_n$ was assigned to a track whose $cos\theta$ was maximum and greater than a specified threshold. The centroid of the track was updated to the centroid of the newly assigned region and its histogram was replaced with the mean of the existing histogram and new region's histogram. Otherwise, if $cos\theta$ was less than the specified threshold, the region was used to initialize a new track.

\item \textbf{Calculate acceleration.} In this step objects' acceleration was computed using the motion descriptors. Average of forward and backward optical flow resulted in acceleration at each pixel $(x,y,t)$ per frame using,
\begin{align*}
g(x,y,t)=F^+(x,y,t)+F^-(x,y,t)
\end{align*}
where $x$ and $y$ are spatial coordinates, $t$ is frame number and $F^+$ and $F^-$ indicate forward and backward optical flow.

The acceleration of regions $ST_i^t$ where $i$ is region index per frame $t$ was computed as the average acceleration of all pixels belonging to that region as:
\begin{align*}
m_i(t)=\frac{1}{|ST_i^t|} \sum \limits_{(x,y) \varepsilon ST_i^t} \left \| g(x,y,t) \right \|
\end{align*}
The resulting acceleration vector was filtered using a Gaussian kernel to remove noise. The result was a motion descriptor of objects in a given input video.

\end{enumerate}

\subsection{Audio Saliency Map Computation}
\label{sec:amc}

For the audio saliency map computation, we used audio-video correlation method proposed in~\cite{li2014s}. The correlation between the aforementioned audio and motion descriptors was used to localize the source of the sound signal in input video frames to indicate audio saliency. Winner-Take-All (WTA) hash~\cite{yagnik2011power}, a subfamily of hashing functions controlled by the number of permutations $N$ and window size $S$, was used to transform both descriptors in rank correlation space. Once in the common rank correlation space, Hamming distance was used to relate the audio signal to an object.

\subsection{Visual Saliency Map Computation}
\label{vsmc}

A classical visual saliency map was used as proposed in~\cite{harel2006graph}. The model approaches the problem of saliency computation by defining Markov chains over feature maps, extracted for features of intensity, color, orientation, flicker, and motion, and treats equilibrium locations as saliency values. In detail, each value of the feature map(s) is considered a node and the connectivity between them is weighted by their dissimilarity. Once a Markov chain is defined on this graph, the equilibrium distribution of this chain computed by repeated multiplication of Markov matrix with an initially uniform vector accumulates mass at highly dissimilar nodes providing activation maps. A similar mass concentration process is applied to these activation maps and output is summed into a final saliency map.

\subsection{Motion Map Computation}
\label{sec:mmc}

Motion map indicates the regions of high motion computed using mean optical flow per frame as described in Section~\ref{sec:visualfeatures}. Adaptive thresholding proposed in~\cite{bradley2007adaptive} was applied on the flows to discard any inconsequential low motion as,
\begin{align*}
I_p=
\begin{cases}
0 & \text{if } I_p < T \cdot I_{avg}\\
1 & \text{otherwise}
\end{cases}
\end{align*}
where pixel $I_p$ is set to zero if its brightness is $T$ percent lower than average brightness of its surrounding pixels.

\subsection{Normalization and Combination}

In this final stage, the three computed maps: a)~visual saliency map, b)~audio saliency map, and c)~motion map were normalized before combining them together into a final audiovisual saliency map. Here the visual saliency map was a sum of normalized activation maps computed using mass concentration algorithm. The other two maps were normalized to a specified range $[0-1]$ using simple linear transformations. The resulting normalized maps were linearly combined to get the final audiovisual saliency map.


\section{Implementation and Evaluation}
\label{sec:mm}


The proposed solution was implemented in MATLAB 2014b and Windows 10 on a 64-bit architecture machine with Intel i5 processor. The same setup was used for evaluation purposes. The parameters used for the proposed solution are given in Table~\ref{tbl:pa}.

\begin{center}
\begin{table}%
\caption{Parameters used for different steps.\label{tbl:pa}}
\centering
\begin{tabularx}{\linewidth}{ p{2.5cm} p{3.5cm} X }
\toprule
Region tracking 	& Search radius ($r$) 			& 100 \\ 
\midrule
\multirow{2}{*}{Audio-video corr.} & No. of permutations ($N$) & 2000 \\
~ 					& Window size ($S$) 		& 5 \\
\midrule
Motion map comp. 	& Threshold \% ($T$)	& 10 \\
\bottomrule
\end{tabularx}
\end{table}
\end{center}

\subsection{Dataset}

Dynamic images and eye movements (DIEM) dataset~\cite{mital2011clustering} was used for evaluation of the proposed approach. The dataset comprises $85$ (eighty-five) videos with or without audio of varying genres. Eye fixation data is collected via binocular eye tracking experiment with 250 participants in total with ages ranging between $18$ and $36$ years with normal/corrected-to-normal vision. In this work, for evaluation $25$ (twenty-five) videos with audio were randomly selected. The video sequences are listed in Table~\ref{tbl:data} along with its properties.


\begin{center}
\begin{table*}%
\caption{Summary of properties of video sequences selected from DIEM dataset. In Audio source column On-screen($+$)/Off-screen($-$): $H=$ human, $N=$ non-human, $M=$ music and $A=$ applause. Properties in order are: Single object($S$)/Multiple objects($M$) ($f_1$), Camera motion ($f_2$), Abrupt scene change ($f_3$), Interaction ($f_4$), Occlusion ($f_5$), Deformation ($f_6$), Crowd ($f_7$), Clutter ($f_8$), and Motion blur ($f_9$). In columns $f_2$ to $f_9$ ($+$) indicates presence and ($-$) indicates absence of the particular property. \label{tbl:data}}
\centering
\begin{tabularx}{\linewidth}{ X p{5cm} p{2cm} p{1.5cm} X X X X X X X X X }
\toprule
No & Video Sequence 	& Scene Type & Audio Source & $f_1$ & $f_2$ & $f_3$ & $f_4$ & $f_5$ & $f_6$ & $f_7$ & $f_8$ & $f_9$ \\



\midrule
 1 & 50\_people\_brooklyn							& Other 			& $H^{+/-}M^{-}$ 		& M & + & + & - & - & - & + & - & + \\
 2 & advert\_bbc4\_bees								& Advertisement 	& $M^{-}N^{+}$ 		& M & - & - & - & - & - & - & - & - \\  
 3 & advert\_bbc4\_library							& Advertisement 	& $M^{-}$ 				& M & - & - & - & - & - & - & + & - \\ 
 4 & advert\_bravia\_paint							& Advertisement 	& $M^{-}N^{+}$ 		& M & - & + & - & - & + & - & - & - \\
 5 & arctic\_bears									& Documentary 		& $H^{-}M^{-}N^{+}$ 	& M & - & - & - & - & - & - & - & - \\
 6 & basketball\_of\_sorts							& Sports			& $M^{-}N^{+}$ 		& M & - & - & + & + & - & - & - & - \\
 7 & BBC\_wildlife\_special\_tiger					& Documentary 		& $H^{-}M^{-}N^{+}$ 	& S & - & - & - & + & - & - & - & - \\
 8 & DIY\_SOS											& Other 			& $H^{+}$ 				& S & - & - & - & - & - & - & - & - \\
 9 & documentary\_adrenaline\_rush					& Documentary 		& $H^{-}M^{-}$ 		& M & + & - & - & + & - & - & - & - \\
10 & documentary\_coral\_reef\_adventure			& Documentary 		& $H^{-}M^{-}N^{+}$ 	& M & + & + & + & + & - & - & - & - \\
11 & game\_trailer\_lego\_indiana\_jones			& Computer Game 	& $H^{-}M^{-}N^{+}$ 	& M & - & + & + & + & + & + & - & - \\
12 & hairy\_bikers\_cabbage							& Other 			& $H^{+}$ 				& M & - & - & + & - & - & - & - & - \\
13 & harry\_potter\_6\_trailer						& Movie 			& $H^{+}M^{-}N^{+}$	& M & - & + & + & + & + & - & - & - \\
14 & home\_movie\_Charlie\_bit\_my\_finger\_again	& Movie 			& $H^{+}$ 				& M & + & - & + & - & - & - & - & - \\
15 & hummingbirds\_closeups							& Documentary 		& $H^{-}N^{+}$ 		& M & - & - & + & - & + & - & - & - \\
16 & music\_trailer\_nine\_inch\_nails				& Crowd 			& $M^{+/-}$ 			& M & - & - & + & + & - & - & - & - \\
17 & news\_bee\_parasites							& News 				& $H^{+/-}$ 			& M & - & - & + & + & - & - & - & - \\
18 & news\_sherry\_drinking\_mice					& News 				& $H^{-}$ 				& M & - & + & + & + & - & - & - & - \\
19 & news\_us\_election\_debate						& News 				& $A^{-}H^{+}$ 		& M & - & - & + & + & - & - & - & - \\
20 & one\_show											& Other 			& $H^{+}$ 				& S & + & - & - & - & - & - & - & - \\
21 & pingpong\_angle\_shot							& Sports			& $N^{+}$ 				& M & - & - & + & - & - & - & - & - \\
22 & planet\_earth\_jungles\_monkeys				& Documentary 		& $H^{-}N^{+}$ 		& M & - & - & - & + & - & - & - & - \\
23 & scottish\_parliament							& Other 			& $H^{+/-}$ 			& M & - & - & - & - & - & + & - & - \\
24 & sport\_football\_best\_goals					& Sports			& $A^{-}M^{-}$ 		& M & + & - & + & + & - & - & - & - \\
25 & stewart\_lee										& Other 			& $H^{+}$ 				& M & - & - & + & + & - & + & - & - \\
\bottomrule
\end{tabularx}
\end{table*}
\end{center}

\subsection{Evaluation Metrics}
\label{sec:em}
The proposed solution was evaluated using four criteria.

\begin{enumerate}
\item \textbf{Area under the curve ($AUC$).} is a location-based metric, where saliency pixels equal to the total recorded fixations are randomly extracted. The true positives ($TP$) and false positives ($FP$) are calculated for different thresholds treating saliency pixels as a classifier. The resulting values are used to plot an ROC curve and compute $AUC$--the ideal score being $1.0$ and a value of $0.5$ indicating random classification.

\item \textbf{Kullback-Leibler divergence ($D_{KL}$).} is a distribution-based dissimilarity measure given as,
\begin{align*}
D_{KL} = \sum_{i} M_f(i) ln \left( \frac{M_f(i)}{M_s(i)} \right)
\end{align*}
it estimates the loss of information when saliency map $M_s$ is used to approximate a fixation map $M_f$--both considered as probability distributions.

The ideal $D_{KL}$ score is zero, meaning the saliency and fixation maps are exactly same, otherwise poorer than the scale of the saliency model.

\item \textbf{Normalized scanpath saliency ($NSS$).} is computed using,
\begin{align*}
NSS = \frac{1}{N} \sum_{i} \frac{M_s(i)-\mu_{M_s}}{\sigma_{M_s}}
\end{align*}
where saliency map $M_s$ is normalized to zero mean and unit standard deviation, then averaged for $N$ fixations. Zero score means a chance prediction whereas a high score indicates high predictability of the saliency model.

\item \textbf{Linear correlation coefficient ($CC$).} is another distribution-based metric computed using,
\begin{align*}
CC = \frac{cov(M_s,M_f)}{\sigma_{M_s} \sigma_{M_f}}
\end{align*}
its output ranges between $-1$ and $+1$, the closer is the score to any of these, the better is predictability of the saliency model.
\end{enumerate}

\subsection{Comparison Methods}
\label{sec:cm}

Based on our literature review, we found no other audiovisual saliency model for complex dynamic scenes. For the sake of comparison, we compared our proposed audiovisual saliency model against two state-of-the-art visual saliency models. The first model proposed in~\cite{liu2014superpixel} derives pixel-level spatial/temporal saliency map from superpixel-level spatial/temporal saliency map constructed using motion and color histogram features. The other spatiotemporal saliency detection model proposed by Liu et al.~\cite{liu2016saliency} is based upon superpixel-level graph and temporal propagation. 


\section{Results}
\label{sec:result}

For evaluation, we computed saliency maps for the selected videos from DIEM dataset using the two state-of-the-art models and our proposed model. Using the evaluation criteria, average scores (Table~\ref{tab:res}) for the resulting saliency maps for the first $300$ frames per video were compared to assess eye movement predictability.

\begin{center}
\begin{table}%
\caption{Average scores for three different techniques on DIEM dataset including our proposed model.\label{tab:res}}
\centering
\begin{tabularx}{\linewidth}{ p{1.0cm} p{1.0cm} p{2.5cm} p{2.5cm} }
\toprule
~ & Ours & Liu et al. 2014~\cite{liu2014superpixel} & Liu et al. 2016~\cite{liu2016saliency} \\
\midrule
$AUC$ 		& \textbf{0.739} & 0.716 & 0.712 \\ 
$D_{KL}$ 	& \textbf{4.153} & 4.255 & 6.437 \\
$NSS$		& 0.913 & 1.091 & \textbf{1.139} \\
$CC$		& 0.147 & \textbf{0.165} & 0.161 \\
\bottomrule
\end{tabularx}
\end{table}
\end{center}


We observe that the proposed model not only outperforms both comparison methods but also results in a satisfactorily higher average score in terms of $AUC$. Moreover, a lower $D_{KL}$ score indicates a better saliency model with less dissimilarity to the ground truth. For the remaining evaluation metrics, $CC$ and $NSS$, the proposed method results in slightly lower scores; however, the results still suggest that the proposed model makes better eye movement predictions, and thus supports the idea of incorporating audio features when computing spatiotemporal saliency for unconstrained videos. Some of the video sequences performed better for instance stewart\_lee, news\_us\_election\_debate and one\_show, with on-screen sound source with no object occlusion, and interaction.



\begin{figure*}[!ht]
\centering
\scriptsize
\bgroup
\def\arraystretch{1.5}
\setlength{\tabcolsep}{2pt}
\begin{tabular}{cccccc}
advert\_bbc4\_bees\_1024x576 &
\includegraphics[width=0.15\linewidth]{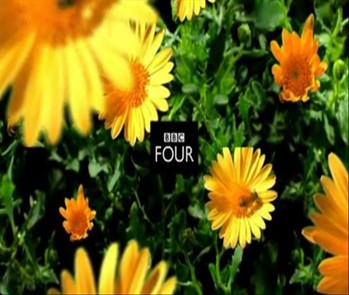} &
\includegraphics[width=0.15\linewidth]{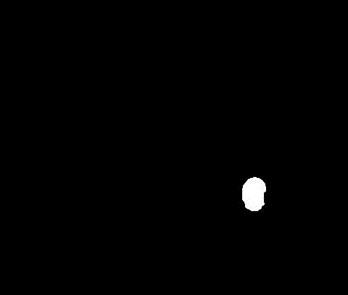} &
\includegraphics[width=0.15\linewidth]{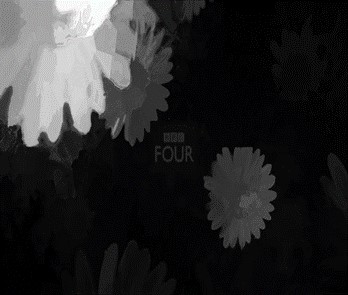} &
\includegraphics[width=0.15\linewidth]{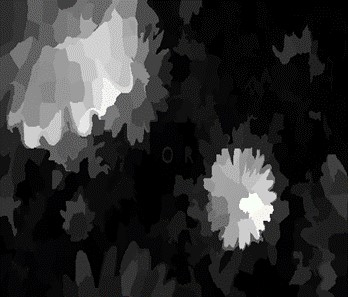} &
\includegraphics[width=0.15\linewidth]{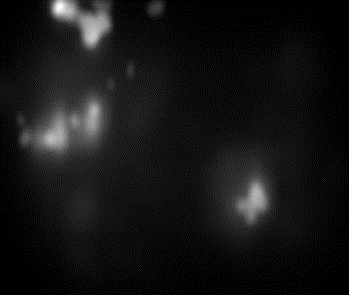} \\

basketball\_of\_sorts\_960x720 &
\includegraphics[width=0.15\linewidth]{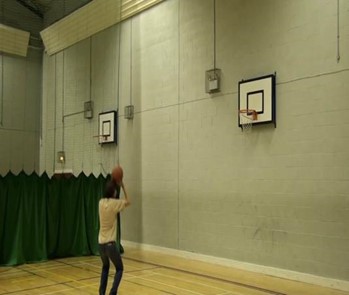} &
\includegraphics[width=0.15\linewidth]{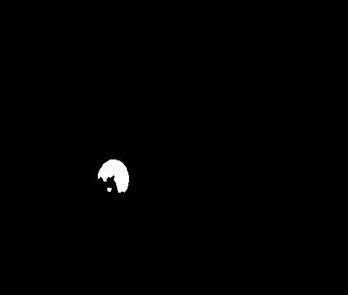} &
\includegraphics[width=0.15\linewidth]{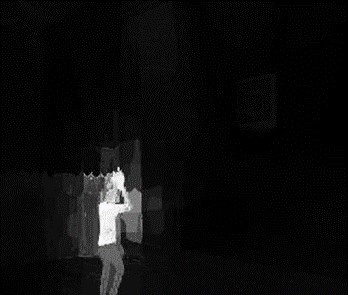} &
\includegraphics[width=0.15\linewidth]{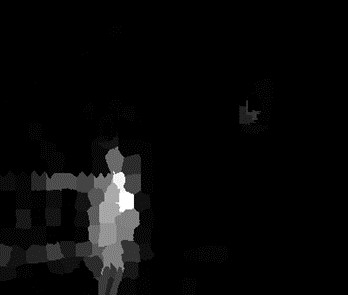} &
\includegraphics[width=0.15\linewidth]{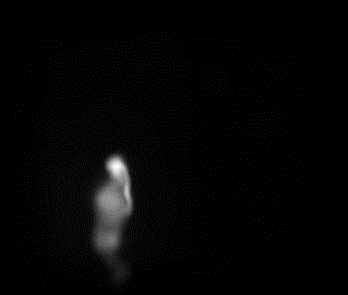} \\

hairy\_bikers\_cabbage\_1280x712 &
\includegraphics[width=0.15\linewidth]{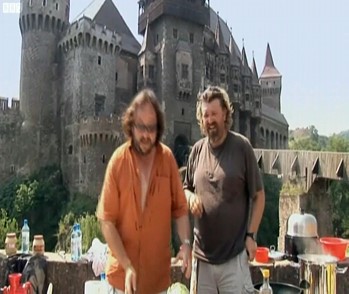} &
\includegraphics[width=0.15\linewidth]{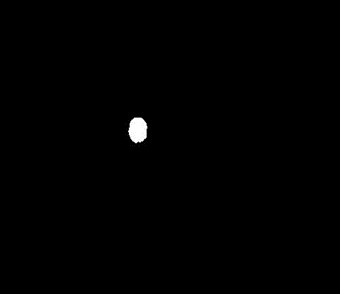} &
\includegraphics[width=0.15\linewidth]{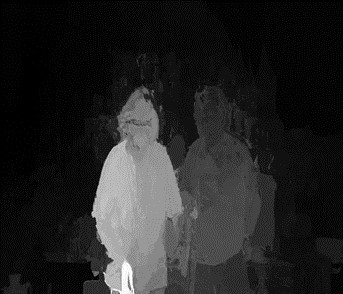} &
\includegraphics[width=0.15\linewidth]{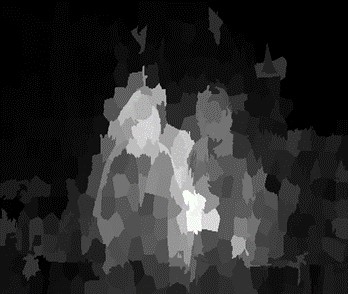} &
\includegraphics[width=0.15\linewidth]{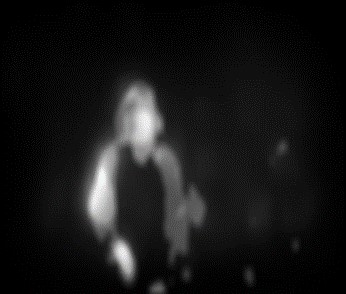} \\

news\_us\_election\_debate\_1080x600 &
\includegraphics[width=0.15\linewidth]{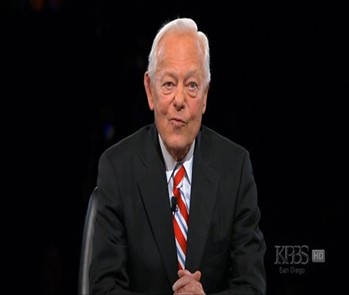} &
\includegraphics[width=0.15\linewidth]{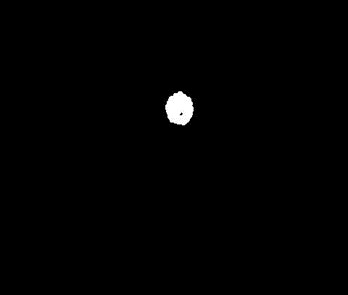} &
\includegraphics[width=0.15\linewidth]{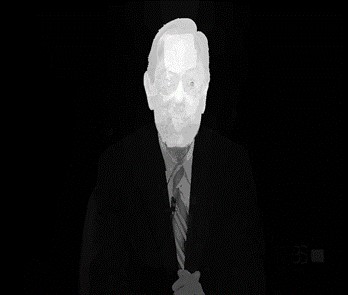} &
\includegraphics[width=0.15\linewidth]{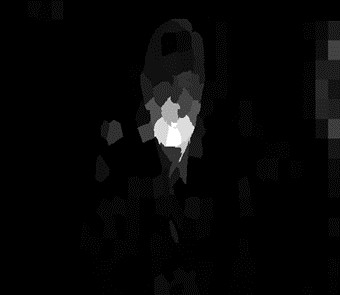} &
\includegraphics[width=0.15\linewidth]{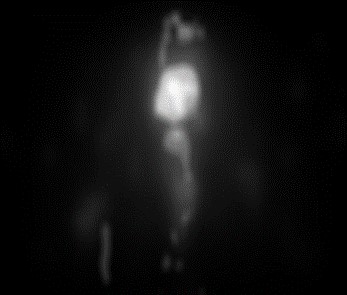} \\

~ & Input & GT & Liu et al. 2014~\cite{liu2014superpixel} & Liu et al. 2016~\cite{liu2016saliency} & Ours \\
\end{tabular}
\egroup
\vspace{5pt}\caption{\label{fig:resultsimg}Comparison of our results with other methods against the ground-truth (GT) on DIEM dataset.}
\label{fig:vr}
\end{figure*}

Figure~\ref{fig:vr} illustrates the saliency maps obtained by different methods. The visual comparison demonstrates that our proposed model performs comparatively better. For instance, video sequence with an on-screen audio source-type in the third row, visual models failed to correspond to the ground truth (GT) as they considered both faces salient; by contrast, the proposed audiovisual model marked the talking face salient.

\section{Discussion}
\label{sec:disc}


Spatiotemporal saliency detection is a challenging problem. It is worth mentioning that existing models ignore the audio signal in the input media. However, a number of experimental studies~\cite{van2008pip,quigley2008audio,song2013different} discuss the influence of aural stimuli on early attention when viewing complex scenes; that is, audio stimuli can provide useful information in guiding eye movements. This influence can be incorporated into existing bottom-up models by the inclusion of low-level audio properties like energy, frequency, amplitude, etc. The resulting audiovisual saliency model makes more sense in application areas like video summarization/compression, event detection, gaze prediction, and robotic vision and interaction. There exist some models in the literature based upon multiple stimuli~\cite{rapantzikos2007audio,coutrot2014audiovisual, ould2016audiovisual} but they lack a generic solution by limiting the models to specific categories of videos. 

A major reason for this lack in literature is due to one of the foremost challenges of audiovisual saliency models: localization of audio source in a given frame. Some methods either use microphone arrays to triangulate a single source or only target stationary sources in a scene. The models fail to perform for dynamic videos, as they assume a single audio source. Furthermore, an approach overcoming these restrictions use correlation analysis between audio and video segments, the audio source is a set of relevant pixels rather than an object. The approach has been used in more recent works where object segmentation precedes audiovisual correlation, making audio source separation maintain the source object shape. Since both audio and video signals are from different domains, reliable correlation requires feature transformation into a suitable space. Moreover, it requires a method to relate an audio descriptor to an object descriptors in a video frame, that is, segmentation and tracking of diversified objects in a video frame is in itself a challenging task. To be precise, the literature lacks in techniques for multiple objects, the case in our dataset with no a priori information about objects like shape, color, size, etc.

In terms of eye movement predictability, the proposed audiovisual saliency model performed better for two evaluation metrics. However, they resulted in comparable scores for the other two metrics. This result can be attributed to the difficulty in segmenting and tracking of multiple interacting objects in varying conditions like motion blur, crowd, etc. Moreover, multiple and/or off-screen audio sources make it a more challenging task to locate an audio source, in consequence, affecting the model's performance.

The proposed saliency model exhibits higher time complexity (Table~\ref{tab:TComp}) attributed to dense optical flow computation, inherently compute-intensive being an optimization problem. The main advantage of using the method is that it estimates both forward and backward flows, and hence the optical flow of occluded regions is also computed correctly. Other alternative motion estimation approaches are block-matching and phase correlation that can be used instead to propose a more efficient solution. Likewise, segmentation of multiple objects is a complex task involving mean-shift segmentation, a non-parametric clustering using kernel density estimation. The approach is not scalable due to its large feature space dimensions. Alternatively, a simpler histogram or superpixel-based segmentation methods can be used to reduce computational complexity, as well as increase model predictability.

\begin{center}
\begin{table}%
\caption{Time complexity for three models including our model for $534 \times 400$ sized video-frame. \label{tab:TComp}}
\centering
\begin{tabularx}{\linewidth}{ p{2.5cm} X p{2.5cm} }
\toprule
Method & \multicolumn{1}{c}{Steps} & time per frame (s) \\
\midrule
\multirow{7}{*}{Ours} & Optical flow & 13.406 \\
~ & Segmentation 		& 15.752 \\
~ & Tracking 			& 0.718 \\
~ & Audio-Video Corr. 	& 2.509 \\
~ & Video Saliency Comp.& 0.218 \\
~ & Motion Map Comp. 	& 0.078 \\
\cmidrule{2-3} 
~ & \multicolumn{2}{c}{32.681s} \\
\midrule
Liu et al. 2014~\cite{liu2014superpixel} & \multicolumn{2}{c}{13.658s} \\ 
Liu et al. 2016~\cite{liu2016saliency} & \multicolumn{2}{c}{7.797s} \\
\bottomrule
\end{tabularx}
\end{table}
\end{center}

A shortcoming of the current study is the use of a subset of the available dataset for evaluation. It may be interesting to perform evaluation using the entire video dataset and/or other available datasets to enforce our finding that aural stimuli alongside visual stimuli can provide useful information in guiding eye movements.

All in all, the proposed solution scored reasonably well, however it can be further improved. An improvement in segmentation and tracking techniques may contribute to a better audio saliency map, and in turn towards a better final saliency map. Furthermore, the use of a more sophisticated visual saliency model, as well as the use of more suitable combination techniques can improve the final result.

\section{Conclusion}
\label{sec:conclusion}

Existing bottom-up saliency models only use visual stimuli while available audio stimuli in the input media remain unused. In this paper, we proposed an audiovisual saliency model incorporating both low-level visual and audio information to produce three different saliency maps: an audio saliency map, a motion saliency map, and a visual saliency map. These maps were linearly combined to get a final saliency map. These maps were evaluated for DIEM dataset using four different criteria. The results show an overall improvement against two state-of-the-art visual saliency models and enforce the idea that of aural stimuli can provide useful information to guide eye movements.

\bibliographystyle{IEEEtran}       
\bibliography{egbib}   

\begin{thebibliography}{10}

\bibitem{avraham2010esaliency}
T.~Avraham and M.~Lindenbaum.
\newblock Esaliency (extended saliency): Meaningful attention using stochastic
  image modeling.
\newblock {\em IEEE Trans. Pattern Anal. Mach. Intell.}, 32(4):693--708, 2010.

\bibitem{birmingham2009saliency}
E.~Birmingham, W.F. Bischof, and A.~Kingstone.
\newblock Saliency does not account for fixations to eyes within social scenes.
\newblock {\em Vision Res.}, 49(24):2992--3000, 2009.

\bibitem{borji2011cost}
A.~Borji, M.N. Ahmadabadi, and B.N. Araabi.
\newblock Cost-sensitive learning of top-down modulation for attentional
  control.
\newblock {\em Mach. Vision Appl.}, 22(1):61--76, 2011.

\bibitem{borji2013state}
A.~Borji and L.~Itti.
\newblock State-of-the-art in visual attention modeling.
\newblock {\em IEEE Trans. Pattern Anal. Mach. Intell.}, 35(1):185--207, 2013.

\bibitem{bradley2007adaptive}
D.~Bradley and G.~Roth.
\newblock Adaptive thresholding using the integral image.
\newblock {\em J. Graphics GPU Game Tools}, 12(2):13--21, 2007.

\bibitem{bruce2005saliency}
N.~Bruce and J.~Tsotsos.
\newblock Saliency based on information maximization.
\newblock In {\em Adv. Neural Inf. Process. Syst.}, pages 155--162, 2005.

\bibitem{cerf2008predicting}
M.~Cerf, J.~Harel, W.~Einh{\"a}user, and C.~Koch.
\newblock Predicting human gaze using low-level saliency combined with face
  detection.
\newblock In {\em Adv. Neural Inf. Process. Syst.}, pages 241--248, 2008.

\bibitem{chang2013superpixel}
H.-S. Chang and Yu-Chiang~F. Wang.
\newblock Superpixel-based large displacement optical flow.
\newblock In {\em 2013 IEEE Int. Conf. Image Process.}, pages 3835--3839. IEEE,
  2013.

\bibitem{coutrot2014audiovisual}
A.~Coutrot and N.~Guyader.
\newblock An audiovisual attention model for natural conversation scenes.
\newblock In {\em 2014 IEEE Int. Conf. Image Process. (ICIP)}, pages
  1100--1104. IEEE, 2014.

\bibitem{coutrot2016multimodal}
A.~Coutrot and N.~Guyader.
\newblock Multimodal saliency models for videos.
\newblock In {\em From Hum. Attention Comput. Attention}, pages 291--304.
  Springer, 2016.

\bibitem{guo2010novel}
C.~Guo and L.~Zhang.
\newblock A novel multiresolution spatiotemporal saliency detection model and
  its applications in image and video compression.
\newblock {\em IEEE Trans. Image Process.}, 19(1):185--198, 2010.

\bibitem{harel2006graph}
J.~Harel, C.~Koch, and P.~Perona.
\newblock Graph-based visual saliency.
\newblock In {\em Adv. Neural Inf. Process. Syst.}, pages 545--552. MIT Press,
  2006.

\bibitem{itti1998model}
L.~Itti, C.~Koch, and E.~Niebur.
\newblock A model of saliency-based visual attention for rapid scene analysis.
\newblock {\em IEEE Trans. Pattern Anal. Mach. Intell.}, 20(11):1254--1259,
  1998.

\bibitem{judd2009learning}
T.~Judd, K.~Ehinger, F.~Durand, and A.~Torralba.
\newblock Learning to predict where humans look.
\newblock In {\em 2009 IEEE 12th Int. Conf. Comput. Vision}, pages 2106--2113.
  IEEE, 2009.

\bibitem{le2007predicting}
O.~Le~Meur, P.~Le~Callet, and D.~Barba.
\newblock Predicting visual fixations on video based on low-level visual
  features.
\newblock {\em Vision Res.}, 47(19):2483--2498, 2007.

\bibitem{le2006coherent}
O.~Le~Meur, P.~Le~Callet, D.~Barba, and D.~Thoreau.
\newblock A coherent computational approach to model the bottom-up visual
  attention.
\newblock {\em IEEE Trans. Pattern Anal. Mach. Intell.}, 28:802--817, 2006.

\bibitem{li2014s}
K.~Li, J.~Ye, and K.A. Hua.
\newblock What's making that sound?
\newblock In {\em Proc. 22nd ACM Int. Conf. Multimedia}, pages 147--156. ACM,
  2014.

\bibitem{liu2016saliency}
Z.~Liu, J.~Li, L.~Ye, G.~Sun, and L.~Shen.
\newblock Saliency detection for unconstrained videos using superpixel-level
  graph and spatiotemporal propagation.
\newblock {\em IEEE Trans. Circuits Syst. Video Technol.}, PP(99):1--1, 2016.

\bibitem{liu2014superpixel}
Z.~Liu, X.~Zhang, S.~Luo, and O.~Le~Meur.
\newblock Superpixel-based spatiotemporal saliency detection.
\newblock {\em IEEE Trans. Circuits Syst. Video Technol.}, 24(9):1522--1540,
  2014.

\bibitem{marat2007video}
S.~Marat, M.~Guironnet, and D.~Pellerin.
\newblock Video summarization using a visual attention model.
\newblock In {\em Signal Process. Conf., 2007 15th Eur.}, pages 1784--1788.
  IEEE, 2007.

\bibitem{marat2009modelling}
S.~Marat, T.H. Phuoc, L.~Granjon, N.~Guyader, D.~Pellerin, and
  A.~Gu{\'e}rin-Dugu{\'e}.
\newblock Modelling spatio-temporal saliency to predict gaze direction for
  short videos.
\newblock {\em Int. J. Comput. Vision}, 82(3):231--243, 2009.

\bibitem{marat2013improving}
S.~Marat, A.~Rahman, D.~Pellerin, N.~Guyader, and Dominique Houzet.
\newblock Improving visual saliency by adding 'face feature map' and 'center
  bias'.
\newblock {\em Cognit. Comput.}, 5(1):63--75, 2013.

\bibitem{mital2011clustering}
P.K. Mital, T.J. Smith, R.L. Hill, and J.M. Henderson.
\newblock Clustering of gaze during dynamic scene viewing is predicted by
  motion.
\newblock {\em Cognit. Comput.}, 3(1):5--24, 2011.

\bibitem{murray2011saliency}
N.~Murray, M.~Vanrell, X.~Otazu, and C.A. Parraga.
\newblock Saliency estimation using a non-parametric low-level vision model.
\newblock In {\em Comput. Vision Pattern Recognit. (CVPR), 2011 IEEE Conf.},
  pages 433--440. IEEE, 2011.

\bibitem{quigley2008audio}
C.~Quigley, S.~Onat, S.~Harding, M.~Cooke, and P.~K{\"o}nig.
\newblock Audio-visual integration during overt visual attention.
\newblock {\em J. Eye Mov. Res.}, 1(2):1--17, 2008.

\bibitem{rapantzikos2007audio}
K.~Rapantzikos, G.~Evangelopoulos, P.~Maragos, and Y.~Avrithis.
\newblock An audio-visual saliency model for movie summarization.
\newblock In {\em Multimedia Signal Process. MMSP 2007. IEEE 9th Workshop},
  pages 320--323. IEEE, 2007.

\bibitem{rapantzikos2009spatiotemporal}
K.~Rapantzikos, N.~Tsapatsoulis, Y.~Avrithis, and S.~Kollias.
\newblock Spatiotemporal saliency for video classification.
\newblock {\em Signal Process.: Image Commun.}, 24(7):557--571, 2009.

\bibitem{ould2016audiovisual}
N.O. Sidaty, M.-C. Larabi, and A.~Saadane.
\newblock An audiovisual saliency model for conferencing and conversation
  videos.
\newblock {\em Electron. Imaging}, 2016(13):1--6, 2016.

\bibitem{song2013different}
G.~Song, D.~Pellerin, and L.~Granjon.
\newblock Different types of sounds influence gaze differently in videos.
\newblock {\em J. Eye Mov. Res.}, 6(4):1--13, 2013.

\bibitem{tatler2011eye}
B.W. Tatler, M.M. Hayhoe, M.F. Land, and D.H. Ballard.
\newblock Eye guidance in natural vision: Reinterpreting salience.
\newblock {\em J. Vision}, 11(5):5--5, 2011.

\bibitem{tavakoli2011fast}
H.R. Tavakoli, E.~Rahtu, and J.~Heikkil{\"a}.
\newblock Fast and efficient saliency detection using sparse sampling and
  kernel density estimation.
\newblock In {\em Scand. Conf. Image Anal.}, pages 666--675. Springer, 2011.

\bibitem{van2008pip}
E.~Van~der Burg, C.N.L. Olivers, A.W. Bronkhorst, and J.~Theeuwes.
\newblock Pip and pop: nonspatial auditory signals improve spatial visual
  search.
\newblock {\em J. Exp. Psychol.: Hum. Percept. Perform.}, 34(5):1053, 2008.

\bibitem{yagnik2011power}
J.~Yagnik, D.~Strelow, D.A. Ross, and R.-S. Lin.
\newblock The power of comparative reasoning.
\newblock In {\em 2011 Int. Conf. Comput. Vision}, pages 2431--2438. IEEE,
  2011.

\end{thebibliography}

%

\end{document}